# IL SANTO SEPOLCRO, ORIENTAMENTO ASTRONOMICO DELLA BASILICA E LE OMELIE DI SAN CIRILLO DI GERUSALEMME

Costantino Sigismondi

Ateneo Pontificio Regina Apostolorum

15 novembre 2012

nel giorno 56° anniversario della laurea a Napoli di mio padre Camillo (20.10.1932-18.10.2011)

**Sommario**

Sulla roccia del Calvario compare sulla scena la Sindone. Su questa roccia si compiono le scritture e questa roccia, posta ad occidente esatto del Tempio, diventa misticamente il nuovo oriente, luogo della Risurrezione del Signore. Gli allineamenti tra i tre luoghi santi alle tre religioni monoteistiche dell'Anastasis, del Tempio/Cupola della Roccia e dell'edicola dell'Ascensione con l'asse Est-Ovest sono qui misurati e commentati sia da un punto di vista astronomico che topografico, alla luce delle nozioni urbanistiche romane di Vitruvio e delle catechesi di san Cirillo vescovo di Gerusalemme 13 anni dopo l'inaugurazione del complesso degli edifici costantiniani al Santo Sepolcro.

**Abstract**

On the spur of Calvary the Shroud appears. On this rock the Scriptures are fulfilled, and this rock, located exactly to the West of the Temple of Jerusalem, becomes mystically the new orient, place of the Resurrection of the Lord. The alignments between the three holy places for the three monotheistic religions, the Anastasis, the Temple, now Dome of the Rock, and the Chapel of the Ascension with the East-West axis are here measured and commented both from an astronomical and topographic point of view, enlighted by the urbanistic roman concepts of Vitruvius and the catechesis of saint Cyril bishop of Jerusalem 13 years after the inauguration of the Constantinian buildings at the Holy Sepulcre.

# 1. Introduzione

La Basilica dell'Anastasis,[1] o del Santo Sepolcro,[2] pur non essendo la chiesa più antica del Mondo, per prenderla a modello per il suo orientamento, e pur non esistendo documentazioni sulla presenza in loco della Sacra Sindone nel primo millennio, è certamente il luogo dove questa ha ricevuto il corpo morto del Salvatore e ce lo ha restituito vivo[3] e dove la Sua immagine è rimasta impressa su quel telo funerario.

Conoscere meglio il Santo Sepolcro ed il suo rapporto con il Tempio, oggi Cupola della roccia, aiuta a capire meglio il mistero del compimento delle Scritture che è li' avvenuto.

San Cirillo di Gerusalemme, vescovo della città, già 13 anni dopo l'inaugurazione della Basilica costantiniana dell'Anastasis vi predicava le sue catechesi. In esse troviamo eco della cultura e mentalità del tempo, trasmesse da un testimone oculare dei lavori di costruzione della Basilica voluta dal pio imperatore Costantino.

Il tema dell'orientamento di un edificio di culto ha un aspetto duplice.

Il primo è teorico, sulle ragioni per cui si debba realizzare un determinato orientamento, e se ne esaminano i testi dalla Sacra Scrittura e dalle catechesi di Cirillo.

Il secondo è pratico, sulla effettiva realizzazione delle prescrizioni religiose o tecniche, e puo' rivelare aspetti d'interesse tecnico scientifico sulle cognizioni topografiche e astronomiche del tempo in cui sono state messe in opera.

Questo studio si inserisce sulla scia di un lavoro sulla misura degli orientamenti[4] delle chiese, dei templi, delle piramidi, e in particolare delle grandi meridiane a foro stenopeico che a partire dal 1475 in Santa Maria del Fiore a Firenze fino a tutto il XIX secolo furono realizzate all'interno di cattedrali e basiliche italiane e francesi.

---

[1] E' il nome greco, storico.
[2] Nome adoperato correntemente, specialmente in lingua inglese. Gli ebrei la chiamano la chiesa della tomba.
[3] «La croce ha ricevuto Gesù vivo e ce lo ha restituito morto; la sindone ha ricevuto Gesù morto e ce lo ha restituito vivo» Beato Sebastiano Valfré, *Alcune notizie concernenti la historia della SS. Sindone con qualche divota agionta indottiva alla Divozione verso la medesima dedicate alle Serenissime Principesse Maria Adelaide e Maria Louisa*, Torino, (1692). cfr. Zaccone, G. M., *Una composizione del beato Sebastiano Valfre sulla sindone*, Centro studi piemontesi, Torino, p. 379-386 (1984).
[4] C. Sigismondi, *Effemeridi,* Ateneo Pontificio Regina Apostolorum (2008).

Si intende verificare quanto la capacità tecnica del tempo, una volta accertata la realizzabilità dell'orientamento ideale Est-Ovest, permettesse la realizzazione di tale allineamento.

Per questi scopi si esamina l'aspetto attuale della basilica dell'Anastasis (§ 2) dopo gli interventi medievali che hanno eliminato quasi del tutto le navate ed il portico ad oriente; la pianta originaria (§ 3) dagli studi del padre Bellarmino Bagatti; i metodi usati per determinare l'orientamento di una parete esterna, esposta al Sole, di un edificio e l'allineamento Calvario-Santo Sepolcro e Santo Sepolcro, Dome of the Rock ed edicola dell'Ascensione sul Monte degli Ulivi, valutando anche la tolleranza di queste misure (§ 4); il caso dell'orientamento della cattedrale di San Giovanni in Laterano, con la lettura biblica della festa estesa a tutta la cattolicità (§ 5); i testi inerenti all'orientamento dalle catechesi di San Cirillo di Gerusalemme (§ 6); seguono, infine, le conclusioni (§ 7).

## 2. La pianta attuale del complesso del Santo Sepolcro

Documentatissima per la sua centralità nel cristianesimo è la basilica dell'Anastasis, e non mi dilungo a descrivere lo stato attuale dell'edificio dove sussiste oggi un difficile condominio tra le varie confessioni religiose, sorvegliato dalle autorità musulmane locali, che rende difficile ogni restauro. Vige la legge dello *statu(s) quo*.[5] Essendo la basilica situata in zona sismica, pero', ogni tanto gli interventi nelle parti comuni sono stati dettati dall'urgenza di metterla in sicurezza come è accaduto proprio per l'edicola del Santo Sepolcro -imbragata in una struttura di piloni di metallo dal 1949- e per la soprastante cupola dell'Anastasis.

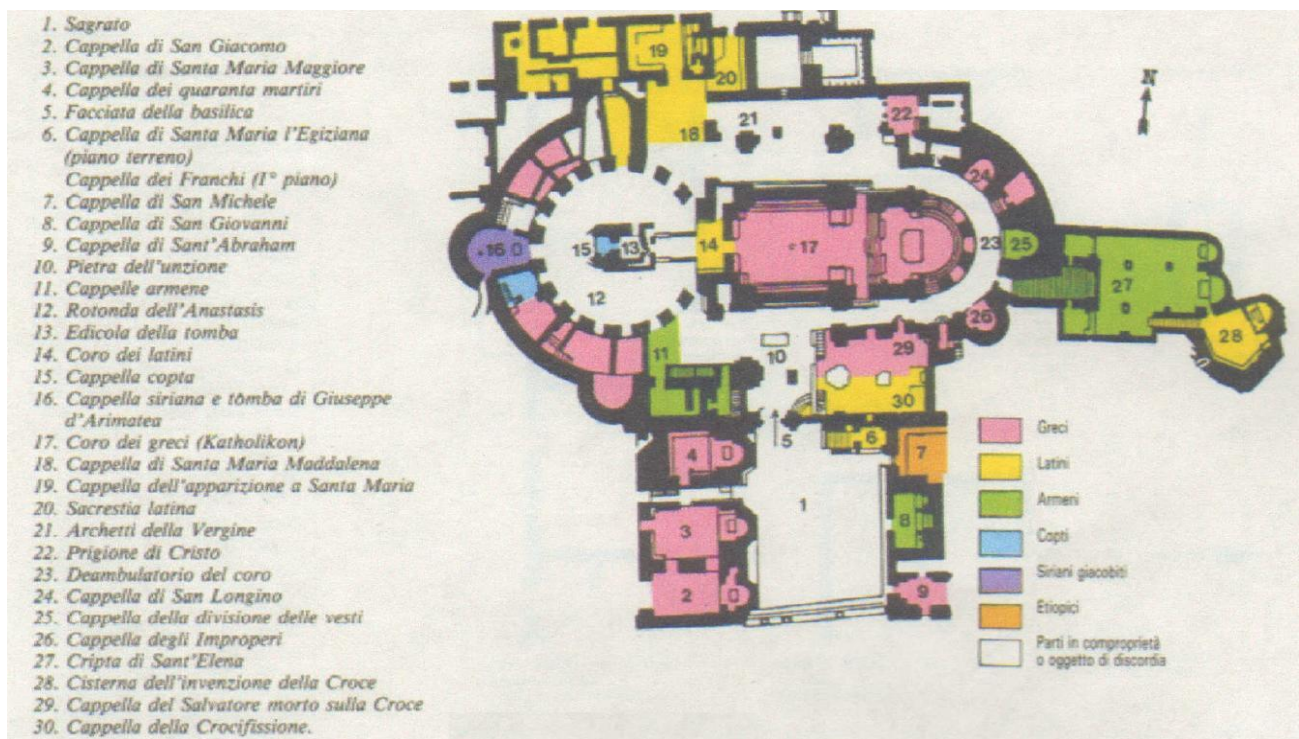

**Figura 1** Il Santo Sepolcro oggi.

---

5   C. Baratto, Guida di Terrasanta, Edizioni Terra Santa, Milano-Jerusalem (1999), si vedano in particolare le zone in bianco sulla mappa, segnalate come «parti in comproprietà o oggetto di discordia»

## 3. Pianta antica del complesso Costantiniano al Santo Sepolcro

Anche qui mi riferisco agli studi storico-archeologici di padre Bellarmino Bagatti OFM,[6] rimandando a quell'articolo per un riferimento puntuale.

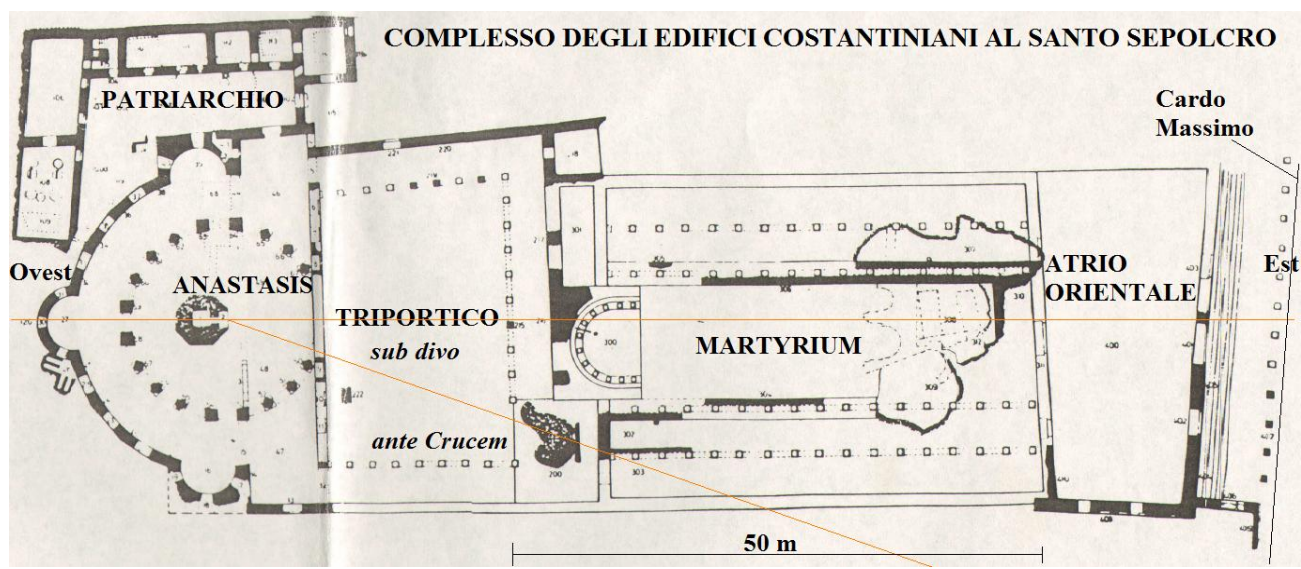

**Figura 2**, adattata da Bagatti (1984).[6]

L'asse del complesso è Est-Ovest, sebbene sia stato necessario un raccordo tra l'atrio orientale ed il cardo massimo, che devia già esso dalla meridiana. L'allineamento tra l'edicola del Santo Sepolcro ed il Calvario indicato in figura forma un angolo di 19° con l'asse Est-Ovest.

La città romana era stata parzialmente ricostruita sotto l'imperatore Adriano che volle rifondarla dopo la vittoria della seconda guerra giudaica nel 135 con il nome di Aelia Capitolina. Effigi per il culto a Giove sul luogo della Risurrezione e ad Afrodite sul Calvario furono installati «da uomini empi»[7] esattamente dove già i Cristiani ricordavano la Passione, Morte e Risurrezione del Signore: questo con l'idea che le religioni siano foriere prima o poi di guerre[8] e quindi era meglio evitare ogni attecchire di nuovi culti.

I lavori di scavo e sbancamento sia della collina del Golgota sia di quella antistante

---

6 B. Bagatti, *La configurazione semiariana delle costruzioni costantiniane del Santo Sepolcro*, Augustinianum 24, 561-571 (1984).
7 Cosi' Eusebio di Cesarea, *De Vita Constantini* **26**, che si guardo' bene di attribuire ad Adriano, predecessore di Costantino, questo epiteto. Cfr. Bagatti op. cit.
8 Idea che è un *leit motif* dei moderni agnostici, che sfuggono le religioni per la loro tendenza a dividere gli uomini dove era la tomba di Gesù hanno modificato anche l'orografia locale.[9]

I nuovi edifici furono stabiliti cosi': la basilica sul sepolcro di Cristo; un'altra posta di fronte sul fianco di est, ma con l'abside voltata verso il sepolcro di Cristo, destinata allo svolgimento liturgico della zona, e fra le due il Calvario rimasto isolato in mezzo ad un chiostro sub divo. Nella mente dei costruttori il Calvario non aveva avuto l'onore di essere «adornato» con un edificio e cio' recava come conseguenza di non essere adatto per compiervi delle funzioni liturgiche.[10] Quelle poche che si riferivano al luogo si tenevano ad una certa distanza, ai piedi, nelle due località che prendono il nome di «ante crucem» e «post crucem» […] rispetto alla chiesa costruita sulla tomba del Signore. Il Calvario ne diveniva una dipendenza secondaria. Le funzioni liturgiche proprie del posto furono due: l'adorazione della croce e le tre ore di agonia che si tenevano nel Venerdi' santo. L'adorazione della croce si teneva la mattina nella cappelleta costruita nel «post crucem» ossia a est del Calvario. Siccome lo spazio era piccolo i fedeli andavano a baciare il legno della croce uno per volta. Quindi si faceva l'oblazione davanti ad un ristrettissimo numero di persone. Le tre ore si tenevano all'«ante crucem» ossia nell'atrio *sub divo*,[11] piovesse o no.[12]

---

9  Al tempo di san Cirillo si discuteva dove fosse stata la tomba del Signore, come si evince dalle sue catechesi:
   Siamo alla ricerca di notizie esatte sul luogo dove fu sepolto. Fu una tomba costruita dalle mani dell'uomo, forse come quelle dei re elevate sul piano della terra? Fu un sepolcro monumentale fatto con pietre collegate insieme, forse coronate nella parte superiore da un fastigio? E quale? Datecela voi la descrizione esatta di questo sepolcro, o profeti. Diteci in quale sepolcro fu deposto il suo corpo e come possiamo cercarlo.
   Rispondono: Guardate la solida roccia che avete tagliato, guardate e vedete (Is 51, 1). Corrisponde a quella di cui leggete nel Vangelo: nella tomba tagliata nella roccia (cf. Lc 23, 53; Mc 15, 46). Come era? Che apertura aveva? Ci risponde un altro profeta: Mi hanno fatto perire chiuso in una fossa e han gettato su di me una pietra (cf. Lam 3, 53); han deposto dentro una pietra per breve tempo me, eletta e preziosa Pietra angolare ma d'inciampo per i giudei e roccia di salvezza per i credenti (1Pt 2, 6. 8; Is 28, 16); hanno piantato me, Albero della vita, nella terra perché essa, maledetta, fosse benedetta per mezzo mio e ne venissero liberati i morti».
10 La regia dei semiariani ha posto il Calvario in posizione defilata, ma le catechesi di Cirillo lo riportano al centro: «Non vergognarti dunque del Crocifisso, ma assieme al profeta confessa con fiducia: Egli si è caricato dei nostri peccati, li piange per noi, e ci ha guarito con le sue piaghe (Is. 53, 4-5). Non siamo ingrati verso il nostro benefattore, e ricordiamoci di quel che segue: Per le iniquità del mio popolo egli è stato messo a morte, vendicherò contro gli empi la sua tomba, e contro i ricchi la sua morte (Is 53, 8-9). Questo proclamò Paolo a chiare note: Cristo è morto per i nostri peccati secondo le Scritture, fu sepolto e il terzo giorno risuscitò secondo le Scritture (1Cor 15, 3-4).»
11 locuz. lat. (propr. «sotto il cielo», «a cielo sereno»), usata in ital. come avv. – Espressione talvolta ripetuta anche in contesti italiani per reminiscenza classica (in questa espressione, che si considera una abbreviazione di quella originaria sub divo coelo, l'agg. divus conserva il valore primitivo di «luminoso»). Nell'archeologia cristiana è chiamato sepolcro sub divo (o anche a cielo scoperto) il sepolcro posto all'aperto, distinto dal sepolcro che si trova dentro una catacomba.
12 Bagatti, op. cit.

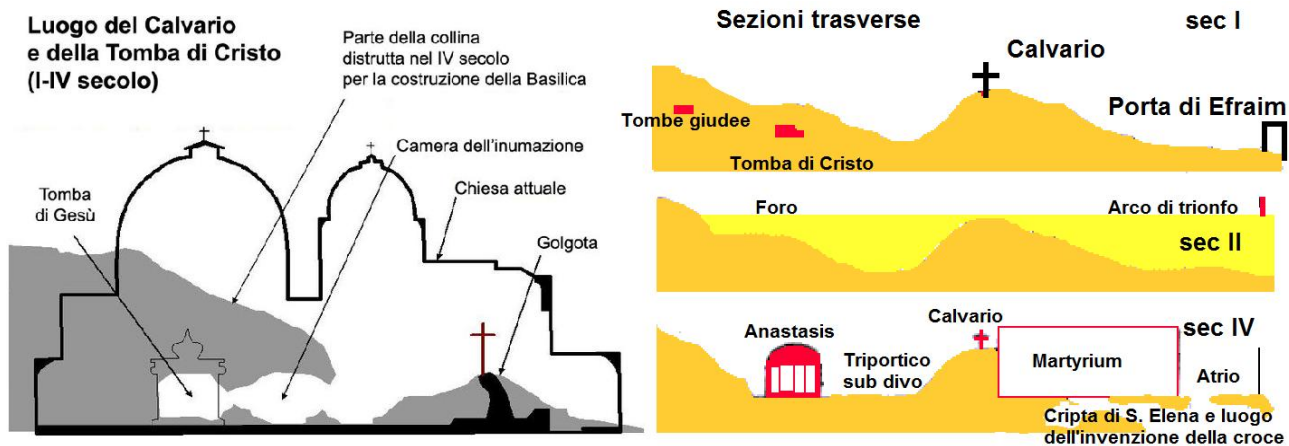

**Figura 3-4** composta da figure riadattate da due fonti web diverse.[13]

Dopo altre vicende con incendi e saccheggi, il 18 ottobre 1009 la basilica costantiniana del Santo Sepolcro fu rasa al suolo dall'Imam/Califfo fatimide al-Hakim bi-Amr Allah, fino alle fondamenta. L'edicola, i muri orientale e occidentale e il tetto della tomba scavata nella roccia in essa racchiusa vennero distrutti o danneggiati (i resoconti dell'epoca sono discordanti),[14] ma i muri nord e sud vennero probabilmente protetti da ulteriori danni dalle macerie. Nel 1048 l'imperatore Costantino IX fece erigere delle piccole cappelle alle rigide condizioni dell'Imam fatimide. I siti ricostruiti vennero conquistati dai cavalieri della prima crociata il 15 luglio 1099.

La prima crociata venne raffigurata come un «pellegrinaggio armato», per cui nessun crociato poteva considerare completo il viaggio senza aver pregato come pellegrino sul Santo Sepolcro. Il capo dei crociati, Goffredo di Buglione, che divenne il primo monarca crociato di Gerusalemme, decise che non avrebbe usato il titolo di "re", e si dichiarò Advocatus Sancti Sepulchri, "Protettore (o Difensore) del Santo Sepolcro". Il cronista Guglielmo di Tiro, colloca la data della ricostruzione a metà del XII secolo, quando i crociati iniziarono a restaurare la chiesa in stile romanico e vi affiancarono un campanile. Questi rinnovamenti unificarono i luoghi santi e vennero completati durante il regno della regina Melisenda, nel 1149.

---

13 http://chemins.eklesia.fr/terresainte/israel/reconst.php e http://it.wikipedia.org/wiki/Basilica_del_Santo_Sepolcro
14 mentre la basilica venne in gran parte demolita, ben tre fonti dell'epoca: una araba Yahya nella sua Storia e due cristiane, Ademaro nel suo Chronicon e Rodolfo il Glabro in Storia, testimoniano, in maniera indipendente ma compatibile tra di loro, che la tomba scavata nella roccia non fu completamente distrutta ma che molte parti di essa sopravvissero.
Iniziamo con Yahya, egli scrive che il figlio di Yaruk e due associati "si impossessarono di tutto l'arredamento che vi si trovava e rasero al suolo la chiesa, tranne ciò che era impossibile distruggere e difficile da estrarre e portare via".
Ademaro di Chabannes, che scrive ad Angouleme nel 1028/1029, riporta nel Chronicon le parole di Raoul de Couhé, vescovo di Périgueux, al suo ritorno da Gerusalemme nel 1010: "quando non riuscirono in nessun modo a ridurre in macerie la tomba, ricorsero a un gran fuoco, ma essa, come un diamante, rimaneva immobile e intatta".
Rodolfo il Glabro fornisce un secondo e indipendente resoconto occidentale della distruzione, derivato forse da Ulrico, vescovo di Orléans (1021-1035), che era a Gerusalemme probabilmente nel 1027. Secondo il suo racconto gli inviati di al-Hakim usarono mazze di ferro per demolire la struttura cava della tomba, ma fallirono nel loro tentativo. Morea, F., *Il Sepolcro di Gerusalemme, la Tomba vuota del Cristo*,
http://www.oessg-lgimt.it/OESSG/terrasanta/SantoSepolcroGerusalemmetombavuotaCristoFabrizioMorea.htm

La chiesa divenne sede dei primi patriarchi latini e fu anche sede dello scriptorium del regno.[15]

## 4. Orientamento del Santo Sepolcro

Il metodo diretto seguito per misurare l'orientamento della Basilica del Santo Sepolcro, o meglio di uno dei muri della facciata, è quello del Sole radente, descritto nel testo Effemeridi.[16] Questo metodo consente di rilevare la direzione di un muro con una precisione migliore del minuto d'arco. Ce n'è un altro applicabile anche a distanza, sulle immagini da satellite, che vado a descrivere: si usa *google maps* e alla massima risoluzione possibile (1 pixel = 50 cm) ho realizzato un collage di 4 immagini allo stesso ingrandimento, che spazia da Ovest (sinistra) a Est (destra) dalle cupole del Santo Sepolcro fino all'edicola dell'Ascensione, passando per la spianata del Tempio. Questo metodo consente precisioni dell'ordine del minuto d'arco, ma su una distanza di circa 1.5 Km.

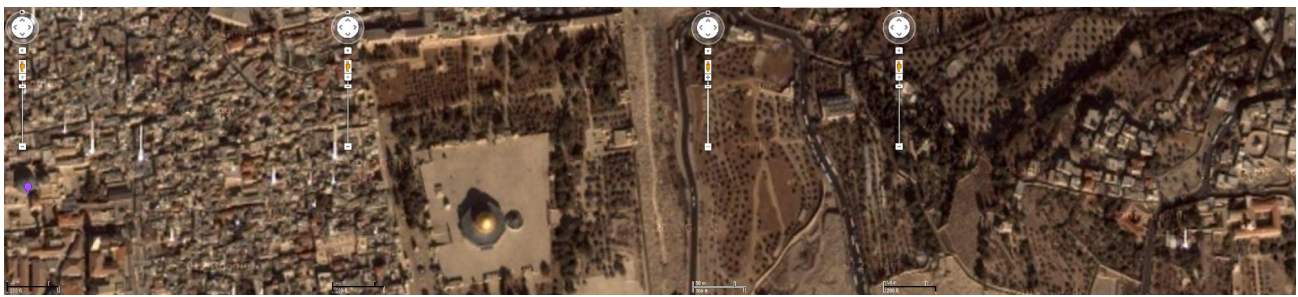

**Figura 5**, fotocomposizione con risoluzione 1 pixel≈50 cm (50m=105 pixel) di immagini di Gerusalemme da satellite (Google maps 2012).

Si lavora direttamente sull'immagine tramite Paint, accessorio di Windows segnando le coordianate in pixel[17] della cupola dell'Anastasis (50,401), della Cupola della roccia (1127,504) e quelle dell'edicola dell'Ascensione (2932,325). Per l'orientamento si considera che l'immagine da satellite è già orientata Est-Ovest, per cui la differenza in pixel della seconda coordinata fornisce direttamente lo spostamento Nord-Sud. Dato anche il collage di immagini ad ogni differenza si puo' associare un'incertezza

---

15 http://it.wikipedia.org/wiki/Basilica_del_Santo_Sepolcro
16 C. Sigismondi, *Effemeridi,* Ateneo Pontificio Regina Apostolorum (2008).
17 Il punto 0,0 è in alto a sinistra

complessiva di ±2 pixel, pari a circa un metro nella scala scelta.

Risultano i seguenti scostamenti dalla linea Est-Ovest:

Santo Sepolcro-Ascensione 1°30'±2' Santo Sepolcro – Cupola della roccia 5°00'±5'

Abbiamo già visto l'allineamento Santo Sepolcro – Calvario pari a 19°, misurato col goniometro sulla pianta in figura 2.

Nell'ipotesi che nessuno dei tre sia casuale, mi appresto a seguire il metodo ermeneutico di san Cirillo di Gerusalemme, ossia di provare con l'antico testamento tutte le testimonianze archeologiche davanti ai suoi occhi. Non è proprio un metodo scientifico come oggi lo si intenderebbe, ma fornirà delle considerazioni interessanti.

## 5. La basilica Lateranense, *Omnium Ecclesiarum Mater et Caput*, tempio rivolto ad oriente? Il metodo di Vitruvio per l'orientamento.

Costantino nel 312[18] fece costruire al Laterano, la *Basilica Salvatoris*, su preesistenze di proprietà imperiale. L'asse maggiore della chiesa è orientato con una rotazione di 12°30' dal punto cardinale Est, verso Nord. Questa direzione, oltre alla probabile influenza delle fondazioni preesistenti, coincide con quella del sorgere del Sole il 12/13 aprile. Si noti anche che la Pasqua del 313 cadde proprio il 13 aprile.[19]

Il riferimento al Laterano, in questa tesi, non è per sostenere l'ipotesi di allineamento astronomico, quanto per considerare la lettura di Ezechiele 47, 1-2, propria della festa della dedicazione della cattedrale lateranense

In quei giorni, [un uomo, il cui aspetto era come di bronzo,] mi condusse all'ingresso del tempio e vidi che sotto la soglia del tempio usciva acqua verso oriente, poiché la facciata del tempio era verso oriente. Quell'acqua scendeva sotto il lato destro del tempio, dalla parte meridionale dell'altare. Mi condusse fuori dalla porta settentrionale e mi fece girare all'esterno, fino alla porta esterna rivolta a oriente, e vidi che l'acqua scaturiva dal lato destro.

La facciata del tempio è rivolta ad oriente, e cosi' dovrebbe essere per ogni chiesa.

---

18 La dedicazione della nuova basilica viene fissata da una tradizione medievale al 9 novembre 312, data in realtà troppo prossima alla sconfitta di Massenzio (28 ottobre); dovendosi tuttavia trattare di una domenica, giorno in genere deputato a tali cerimonie, si potrebbe ipotizzare sulla base del calendario il 318, domenica 9 novembre. Le Chiese Paleocristiane di Roma, Roma Archeologica n. 16-17, marzo 2003, Elio de Rosa Editore, p. 38.
19 Sigismondi, C., *Effemeridi*, p. 69-70 (2008).

Le deviazioni macroscopiche dall'oriente sono dovute al fatto che la chiesa è stata edificata su un tempio preesistente, mentre le deviazioni meno importanti, come in San Pietro in Vaticano, dell'ordine del grado, sono legate alla precisione con cui le tecniche di orientamento venivano applicate dagli architetti dell'epoca. Talvolta, come nel caso delle piramidi di Gizah,[20] e probabilmente anche del Pantheon,[21] erano le stesse tecniche ad includere un errore sistematico, ed anche la loro applicazione scrupolosa ha indotto uno spostamento sistematico nell'allineamento.

Tutto questo per dire che lo studio dell'allineamento degli edifici di culto permette anche di conoscere il grado di tecnologia topografica dell'epoca della costruzione.

Già l'orientamento dei *castra* e delle città romane, per essere protetto dai venti dominanti, come si legge in Vitruvio,[22] avveniva seguendo una tecnica di individuazione della linea meridiana applicata dai *gromatici* e basata sull'ombra descritta da uno stilo-gnomone su un piano.

Sul piano si disegna un cerchio centrato sulla base dello gnomone e passante per l'estremo dell'ombra attorno all'ora quinta. Le due intersezioni dell'ombra con il cerchio rispettivamente prima e dopo il mezzodì descrivono un segmento. La linea meridiana parte dalla base dello gnomone e biseca questo segmento.[23]

---

20 Spence, K., *Ancient Egyptian chronology and the astronomical orientation of pyramids,* Nature 408, 324 (2000).
21 Sigismondi, C., *Effemeridi*, p. 48-49 (2008).
22 L'architetto secondo Vitruvio «Astrologiam caelique rationes cognitas habeat.» *De Architectura* **I**, 1, 3. Vitruvio tra il 35 ed il 25 avanti Cristo raccolse nel suo trattato la sintesi di tutto il sapere teorico e pratico acquisito negli ultimi due secoli ellenistici nel campo dell'architettura e dell'ingegneria.
23 «ut inveniantur regiones et ortus eorum [ventorum] sic erit ratiocinandum.
**6.** conlocetur ad libellam marmoreum amussium mediis moenibus, aut locus ita expoliatur ad regulam et libellam ut amussium non desideretur, supraque eius loci centrum medium conlocetur aeneus gnomon indagator umbrae, qui graece σκιαθήρας dicitur. huius antemeridiana circiter hora quinta sumenda est extrema gnomonis umbra et puncto signanda, deinde circino diducto ad punctum quod est gnomonis umbrae longitudinis signum, ex eo a centro circumagenda linea rotundationis. itemque observanda postmeridiana istius gnomonis crescens umbra, et cum tetigerit circinationis lineam et fecerit parem antemeridianae umbrae postmeridianam, signanda puncto.
**7.** ex his duobus signis circino decussatim describendum, et per decussationem et medium centrum linea perducenda ad extremum, ut habeatur meridiana et septentrionalis regio.» Vitruvio *De Architectura* **I,** 6, 5-7

# 6. L'orientamento dei luoghi santi nelle catechesi di san Cirillo di Gerusalemme

Cirillo, vescovo di Gerusalemme, scrisse le catechesi nel 348, 13 anni dopo l'inaugurazione del complesso costantiniano del Santo Sepolcro. Era stato testimone dell'edificazione di quegli edifici, e conosceva bene il sostrato culturale e religioso del tempo.[24] L'approccio catechetico del presbitero di Gerusalemme è spesso confortato dalla constatazione veritativa e significativa dei luoghi santi.[25]

I punti cardinali connessi con il Sole hanno un simbolismo spirituale.

L'occidente è il luogo delle tenebre, di satana.

Appena entrati nel vestibolo dell'edificio dove si amministra il battesimo, standovene rivolti in piedi verso Occidente, avete ascoltato l'ordine di stendere la mano e di rinunziare a satana come se fosse presente. [Cat. XIX, 2]

E ancora specifica il motivo del rivolgersi ad occidente:

Ma risuona ancora alle tue orecchie l'ordine di stendere la mano dicendo al demonio come a un vicino cui si parli: «Rinunzio a te satana».
Voglio ora dirvi perché vi siete volti all'Occidente, è necessario spiegarlo: siccome l'Occidente è la regione materiale delle tenebre, e il demonio è oscurità che domina nelle tenebre, avete guardato a Occidente per rinunziare con gesto simbolico al principe delle tenebre e delle caligini. [Cat. XIX, 4]

L'oriente è il luogo del paradiso, della luce, della professione di fede.

Con la rinunzia a satana, definitivamente sciolta ogni alleanza con lui e rotti gli antichi patti con l'inferno (cf Is 28, 15), eccoti schiuso il paradiso che Dio piantò ad Oriente (cf Gen 2, 8) e da cui fu cacciato il nostro progenitore caduto nella trasgressione (cf Gen 3, 25).

---

24 Non è questo il luogo adatto ad una presentazione esauriente di questo grande padre della Chiesa d'Oriente che visse tra il 315 e il 387. Pastore instancabile esortava i cristiani a conoscere bene le dottrine degli avversari per non restarne vittima a loro volta. Dimostra ogni sua affermazione ricorrendo alla Santa Scrittura, mostrando come sempre l'Antico Testamento si inveri infallibilmente nel Nuovo, per confutare e confondere ebrei ed eretici. Partecipo' al secondo concilio di Costantinopoli, riconosciuto come degno vescovo della chiesa madre di tutte le chiese, dopo aver sofferto vari anni di esilio. Leone XIII nel 1882 lo dichiarò Dottore della Chiesa.
25 Riggi, C., *Cirillo di Gerusalemme, le Catechesi*, Città Nuova (1993) in corso di revisione da G. Berbenni.

Questo è il significato del gesto che fai di volgerti dall'Occidente all'Oriente, regione della luce, e della professione di fede che ti si richiede di fare nello stesso momento dicendo: Credo nel Padre, nel Figlio e nello Spirito Santo, e in un solo battesimo di penitenza. [Cat. XIX, 9]

Ed il luogo della morte in croce, il Golgota, è il centro del Mondo, ricordato anche nel simbolismo della croce di Terrasanta.

Sulla croce allargò le sue mani per abbracciare con il Golgota, posto proprio al centro della terra, tutto il mondo fino ai suoi estremi confini. Non sono io ad affermarlo, ma lo dice il profeta: Hai operato la salvezza dal centro della terra (Sal. 74 / 73, 12). [Cat. XIII, 28]

Per questo dal Golgota si va al Monte degli Ulivi, ad oriente, per l'ascensione.

Avremmo di che arrossire se, una volta crocifisso e posto nel sepolcro, vi fosse rimasto chiuso; ma egli dopo essere stato crocifisso qui sul Golgota, ascese al cielo: dal Monte degli Ulivi, da lì ad Oriente. Così, dopo essere disceso agli inferi e risalito sulla nostra terra, ascese al cielo donde il Padre gli fece sentire la sua voce: «Siedi alla mia destra, finché avrò posto a scanno dei tuoi piedi i tuoi nemici». [Cat. IV, 13]

Ancor oggi il Monte degli Ulivi si erge per mostrare agli occhi dei fedeli chi di lì ascese sulle nubi additando la porta per salire al cielo: colui che disceso dal cielo fino a Betlemme, dal Monte degli Ulivi ascese al cielo; dal cielo disceso fra gli uomini per ingaggiare la lotta, dal Monte degli Ulivi ascese per riceverne la corona. Tra le tante testimonianze ritieni anche queste del luogo dove egli risorse e quelle ad oriente donde ascese: queste dove gli angeli lo annunziarono, e quelle donde egli salì sulla nube e i discepoli discesero. [Cat. XIV, 23]

Segno del re che doveva venire fu certo anche il fatto che cavalcò un asinello; ma indicaci quello più grande, il segno del luogo dove il re porrà termine al cammino iniziato con quell'ingresso.
Non dovrai indicarci un luogo da noi sconosciuto, lontano dalla città, mostraci un segno qui vicino, accessibile ai nostri occhi, che possiamo contemplare dentro la città.
Il profeta ce ne dà l'indicazione: «In quei giorni i suoi piedi si poseranno sopra il Monte degli Ulivi che sta di fronte a Gerusalemme verso oriente».
È questo il luogo che può vedere chiunque, anche restando dentro la città. [Cat. XII, 11]

Nel 351 Cirillo vedrà perpetuarsi nella gloria la tragedia del Golgota, all'apparire di «una gigantesca croce di luce estendentesi fino al santo Monte degli Ulivi». Non gli sembrò un'illusione collettiva ma la profezia di «molti anni pacifici».[26]

## 7. Conclusioni: allineamento verso l'Ascensione

L'attuale edicola dell'ascensione è un punto topograficamente rilevante, esattamente sul parallelo passante per la cupola dell'Anastasis. La scelta è in perfetta rispondenza con le catechesi di san Cirillo. Il luogo reale dell'ascensione è ignoto (Lc parla genericamente di «monte degli Ulivi, verso Betania», cfr. Lc 24,44-53 e Atti 1,3-12). Questo luogo ricorda l'Ascensione per i cattolici, mentre gli ortodossi la ricordano all'interno del monastero russo, il luogo dell'incontro tra Paolo VI ed Atenagora. Nella piccola edicola e nel recinto, solo il giorno della festa dell'Ascensione, si alternano i cristiani delle varie confessioni nella celebrazione dei loro riti, l'ufficio divino e la S. Messa, sotto la sorveglianza dei musulmani che la trasformarono in moschea: è l'unico caso al Mondo in cui lo permettono.

In epoca bizantina esisteva in questa zona una chiesa detta "Imbomom"[27], cioè "sulla vetta",[28] fatta costruire da Pomenia, una ricca matrona, nel 376[29] e visitata due anni dopo dalla pellegrina Egeria; la chiesa, incendiata dai persiani, fu restaurata successivamente e visitata nel 670 dal pellegrino Arculfo.

Il recinto, entro cui è racchiusa l'edicola dell'ascensione, sorge sui resti della costruzione crociata del 1152 ed ha la forma di un ottagono. Anche l'edicola, ornata di archetti sostenuti da colonnette con capitelli semplici, è di origine crociata e fu trasformata in moschea nel 1198, dopo la vittoria di Saladino nel 1187.

L'edicola si trovava all'interno della grande chiesa crociata della quale rimangono ancora parte delle mura. La cupola era aperta verso il cielo per un evidente motivo simbolico. Nel 1200 l'edicola fu chiusa in alto ed è giunta così fino a noi. All'interno è venerata da una tradizione cristiana e musulmana (la fede musulmana ammette l'ascensione di Gesù, ma non la sua morte e resurrezione) una pietra, isolata nel pavimento, sulla quale si vuol vedere l'impronta del piede sinistro di Gesù. Il pellegrino Arculfo (VII sec.) narra che la folla si accalcava per raccogliere la polvere sopra le impronte. La tradizione delle impronte di Gesù è dunque molto antica e testimoniata sin dalle lettere di Paolino da Nola (Ep. 31,4, circa il 401: «Così in tutta la superficie della basilica solo questo luogo rimane verdeggiante e la terra offre alla venerazione dei fedeli l'impronta dei piedi del Signore, in modo che davvero si può dire: noi lo abbiamo adorato là dove si sono posati i suoi piedi»).[30]

---

26 Cf. *Sancti Cyrilli epistola ad Constantium piissimum imperatorem. De signo lucidae crucis Hierosolymis viso, quod in caelis apparuit*, PG 33, 1165-1176.
27  Dall'aramaico bâmâ che significa altezza.
28  A 818 metri sul livello del mare.
29  Aviva Bar−Am, *Beyond the walls: churches of Jerusalem*, Ahva Press Jerusalem (1998).

Dalle misure topografiche è chiaro che l'allineamento tra il santo Sepolcro con il Tempio e con l'edicola dell'Ascensione lungo l'asse Est-Ovest non è un caso.

Da un lato ci si puo' chiedere se questo allineamento Sepolcro – Ascensione sia stato voluto e considerare l'aspetto tecnico della riuscita entro 1°30' da datare al 378 con la costruzione voluta dalla matrona Pomenia.

Dall'altro lato l'allineamento con il Tempio del Calvario è un dato di fatto: compie le scritture, per dirla con Cirillo.

In fine i 19° da Ovest verso Nord con cui dal Calvario si osserva il Sepolcro indicano l'azimut dove il Sole tramonto' il Venerdi' Santo? No, l'azimut del tramonto del Sole, fino ad un mese dopo l'equinozio vernale, non arriva a discostarsi dall'Ovest di 15°.[31]

Tutto questo lavoro riguarda la Sindone[32] solo indirettamente, ma tratta di argomenti che Cirillo stesso chiamerebbe senza dubbio a testimoniare della Risurrezione insieme al Sole a alla Luna «un'altra testimonianza della risurrezione dei morti puoi chiaramente prenderla tra i luminari del cielo.» [Cat. XVIII, 10] e ai luoghi stessi:

«Tra le tante testimonianze ritieni anche queste del luogo dove egli risorse e quelle ad oriente donde ascese». [Cat. XIV, 23]

---

30 Cioni, M., *Edicola dell'Ascensione,* http://www.gliscritti.it/gallery3/index.php/album_001/Gerusalemme/Moschea-dell_Ascensione
31 Calcoli eseguiti con le effemeridi di calsky.org e con il programma Ephemvga.
32 Anche Cirillo ne parla esplicitamente una volta sola: «Vera la morte di Cristo, vera la separazione della sua anima dal suo corpo, vera anche la sepoltura del suo santo corpo **avvolto in una sindone** (cf Mt 27, 59). In lui tutto è veramente avvenuto, per voi invece non è avvenuta che una somiglianza della sua morte e della sua passione.» [Cat. XX, 7]

# Bibliografia